\documentclass[showpacs,twocolumn]{revtex4-1}
\pdfoutput=1
\usepackage{amsmath,amssymb}
\usepackage{url}
\usepackage{graphicx}
\usepackage{epstopdf,epsfig,color}
\newtheorem{theorem}{Proposition}[section]

\begin{document}
\title{A stochastic microscopic model for the dynamics of antigenic variation}

\author{Gustavo Guerberoff}
\email{\tt gguerber@fing.edu.uy Corresponding author.}
\affiliation{Instituto de Matem\'atica y Estad\'{\i}stica ``Prof. Ing. Rafael Laguardia'', Facultad de Ingenier\'{\i}a,
Universidad de la Rep\'ublica, Montevideo, Uruguay}

\author{Fernando Alvarez-Valin} 
\email{\tt falvarez@fcien.edu.uy}
\affiliation{Secci\'on Biomatem\'atica,  Facultad de Ciencias, Universidad de la Rep\'ublica, Montevideo, Uruguay}


\begin{abstract}
{\bf Abstract:} We present a novel model that describes the within-host evolutionary dynamics of parasites undergoing antigenic variation. The approach uses a multi-type branching process with two types of entities defined according to their relationship with the immune system: clans of resistant parasitic cells (i.e. groups of cells sharing the same antigen not yet recognized by the immune system)  that may become sensitive, and individual sensitive cells that can acquire a new resistance thus giving rise to the emergence of a new clan. The simplicity of the model allows analytical treatment to determine the subcritical and supercritical regimes in the space of parameters. By incorporating a density-dependent mechanism the model is able to capture additional relevant features observed in experimental data, such as the characteristic parasitemia waves. In summary our approach provides a new general framework to address the dynamics of antigenic variation which can be easily adapted to cope with broader and more complex situations.

\vspace{2 mm}

\end{abstract}

\maketitle

\section{Introduction}

\vspace{4 mm}

Parasites have evolved a diversity of sophisticated strategies to evade the host's immune response, among which antigenic variation is perhaps one of the most striking ones. This strategy consists in periodically changing a protective coat composed by an abundant and immunogenic protein. In this mechanism the parasites express only one variant antigenic protein copy from a large repertoire of silent genes. The mechanism allows transient immune evasion, since after changing the variable protein that is being expressed, an entirely new parasite population arises that is not recognized by the host's immune system that has developed an antibody response directed against the previous antigen. By repeating this cycle during the course of an infection, parasites are able to remain in the host for long periods of time.

Perhaps the most paradigmatic example is that of African trypanosomes (responsible of producing the sleeping sickness in humans), but antigenic variation is also observed in {\it Giardia llambia}, and the malaria agents belonging to the {\it Plasmodium} genus. Some viruses are also able to evade the immune response in a strategy similar to that just described. However in the latter systems antigenic diversity is generated by the introduction of point mutations in the gene encoding the antigen, rather than by switching the expressed gene.  This implies some substantial differences in the dynamics since the new antigen is most likely somewhat similar to the previous one (parental) and perhaps is recognized (yet with lower affinity) by the same antibodies

Several models have been developed to study and predict the population dynamics of parasites and viruses during the course of an infection within a single host. Most of them are based on a  system of coupled differential equations inspired on variations of the predator-prey models (see ref. \cite{Kosinski} to \cite{Blyuss}). In short, this approach consists in a set of differential equations describing the dynamical interaction between antigens and the host's immune system, in such a way that the outcome of one equation is a modulating parameter of the others, and including in some cases cross-reactive immune responses as well as other possible interactions. Stochasticity is incorporated ad hoc into the models by the emergence of new variants (which eventually are  not recognized by immune system) at random times, usually driven by a Poisson process.

Very recently Gurarie et al. (see ref. \cite{Gurarie}) implemented a discrete time computer model for the case of malaria. This modeling approach, termed agent-based, consists in a set of coupled difference equations that describe the transition between successive iterations of the parasite population (i.e. parasite generations) and its interaction with the immune system. According to the authors the advantage of this approach is that, owing to its discrete nature, the aleatory components are incorporated more easily by adding random factors to the variables that represent the efficiency of immune system.  

In spite of the existence of these models of antigenic variation, in our opinion it is worth re-addressing the problem from a different perspective. Here we present a model that tackles this topic from a microscopic point of  view that consists in following the pathway and behavior of its individual elements through a multi-type branching process. Iwasa, Michor and Nowak (ref. \cite{Iwasa}) already used this methodology to study problems related to the ones presented here; however these authors focused into the evolutionary dynamics of viruses to escape antiviral therapy.

The model presented here has the following advantages: the role of each one of its parameters has a straightforward biological interpretation; its versatility easily permits the incorporation of increasing complexity and realism; the process can be studied  backwards in time, like in population genetics' coalescent theory. Finally, its simplicity allowed us to obtain an analytical expression for the critical surface separating subcritical from supercritical regimes in the parameter space.

\section{The model}

Our model, a discrete-time non-independent multi-type branching process, assumes the existence of two types of {\em cells/infective particles} (viral particles, parasites, etc.) which are defined according to the host's immune system ability to recognize them; namely, sensitive (type-1) and resistant (type-2) cells. The model involves three parameters, $\delta, \mu, p \in [0,1]$, that are defined as follows:

\vspace{2 mm}
 
The population of cells proliferates by binary division and the offspring of sensitive cells die, independently, with probability $\delta$ (and consequently survive with probability $1-\delta$). Surviving cells may become resistant (i.e. start producing a new antigen variant) with probability $\mu$.


A newly arisen resistant cell creates a clan (or dynasty) of resistant cells in the following, recursive, way: at a given time the whole progeny of resistant cells divides into resistant cells, which remain as such with probability equal to $p$. In other words, $1-p$ is the probability that the immune system acquires the ability to recognize this particular clan, i.e. the clan bearing this specific variant antigen. 

This means that for a given resistant cell appearing at generation $n$ we take a geometric random variable, $N$, of parameter $1-p$, and consider the whole dividing resistant clan until generation $n + N$, where resistant cells become sensitive.


Note that while the probabilities $\delta$ and $\mu$ are intrinsic properties of parasitic cells, $p$ measures the immune system capability to start recognizing a previously unidentified variant antigen (and consequently a clan).  

Summarizing: parameter $\delta$ measures the efficiency of immune response against sensitive cells; parameter $\mu$ represents the rate at which new resistant variants appear; and parameter $p$ is related to the delay times spent by the immune systems to recognize a new variant.  

\begin{figure}
\includegraphics[width=0.8 \textwidth]{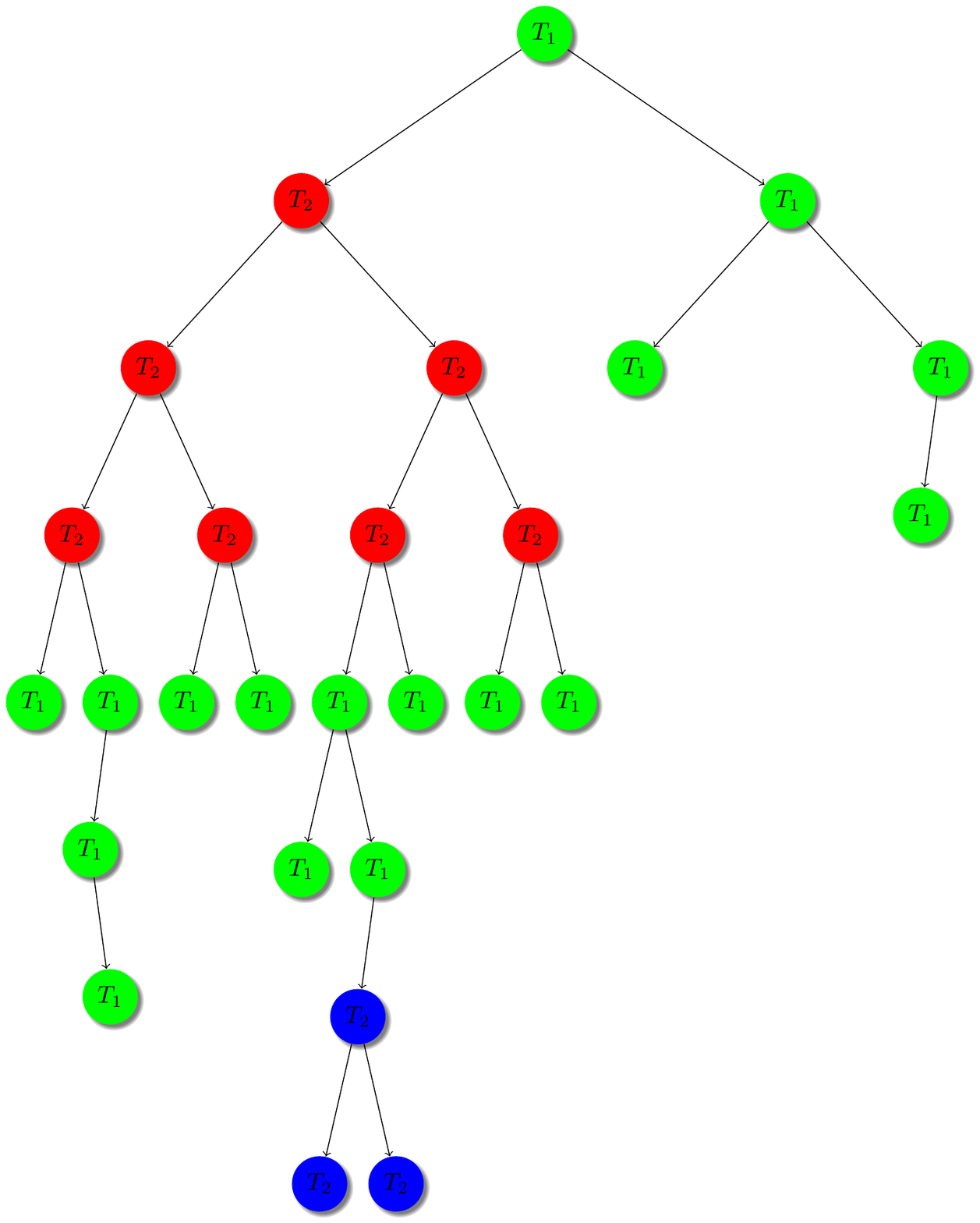}
\caption{Green cells are sensitive. Red and blue cells represent clans of antigen variants not recognized by the immune system.}
\label{fig1}
\end{figure}


Figure \ref{fig1} illustrates a realization of the process: a sensitive (green) cell originates a resistant descendant clan (red) which in turn become sensitive (green) after three generations. At the bottom of the figure, the emergence of a new resistant variant (blue) is represented. Different clans of resistant cells, and sensitive cells, evolve independently.


We remark that this is not an standard multi-type branching process as for example those considered in ref. \cite{Kimmel}, in the sense that resistant cells in a given clan do not evolve independently: instead, their destiny is determined by immune system capacity, which does or does not recognize the whole population of cells carrying a specific variant antigen. The model could also be envisaged as a percolation process on the complete binary tree in presence of a random environment (the clans of resistant cells of random sizes).  

\section{Extinction probability}

To compute the extinction/survival probabilities of the process --and thus obtaining the critical surface as a function of the parameters-- we introduce an additional multi-type branching process which, clearly, has the same extinction/survival probabilities as the antigenic variation model introduced before. This independent multi-type branching process with two types of cells is obtained from the antigenic variation model by {\em collapsing} to one generation each clan of resistant cells.

\subsection{Independent multi-type branching process}

Let consider two types of cells that evolve independently. The progeny of each cell is as follows:

\vspace{2 mm}

Type-1 cells give birth to:

\begin{itemize}
\item two type-1 cells with probability $(1-\mu)^2(1-\delta)^2$,
\item two type-2 cells with probability $\mu^2(1-\delta)^2$,
\item one type-1 cell and one type-2 cell with probability $2 \mu (1-\mu)(1-\delta)^2$,
\item one type-1 cell with probability $2 (1-\mu) \delta (1-\delta)$,
\item one type-2 cell with probability $2 \mu \delta (1-\delta)$,
\item no cell with probability $\delta^2$.
\end{itemize}

\vspace{2 mm}

Type-2 cells give birth to:

\begin{itemize}
\item $2^{N}$ type-1 cells with probability $p^{N-1}(1-p)$,
$N=1,2,3, \ldots$
\end{itemize}

\subsection{Extinction probability}

We get the equation for the extinction probability of this independent multi-type branching process --starting with one type-1 cell, one type-2 cell, or eventually any given initial configuration of cells-- following standard procedures that use probability generating functions (pgf) as given, for example, in ref. \cite{Kimmel}, \cite{Jagers}, \cite{Haccou}, \cite{Serra}.

For each $n=0,1,2, \ldots$ we denote $Z_1(n)$ (resp. $Z_2(n)$) the
number of type-1 (resp. type-2) cells present at generation $n$. In the
particular case $n=1$ we put $Z_1 = Z_1(1)$ (resp. $Z_2 = Z_2(1)$) to simplify the notation.




As it is well known, the distribution of $Z_1(n)$ and $Z_2(n)$ as much as the probability of extinction of the process could be computed from the pgf's of $Z_1$ and $Z_2$,  which are:

\begin{eqnarray}
f_1(s,t) &=& \mathbb{E} \left[ s^{Z_1} \, t^{Z_2} | Z_1(0)
=1, \, Z_2(0) = 0 \right] \nonumber \\
f_2(s,t) &=& \mathbb{E} \left[ s^{Z_1} \, t^{Z_2} | Z_1(0) =0, \,
Z_2(0) = 1 \right], \nonumber
\end{eqnarray}

$s,t \in [0,1]$.

\vspace{4 mm}

We have:

\begin{eqnarray}
f_1(s,t) &=& \left\{ \delta + (1-\delta)[ \mu t + (1-\mu) s ] \right\}^2  \nonumber \\
f_2(s,t) &=& (1-p) \sum_{k=1}^{\infty} s^{2^k} p^{k-1}. \nonumber
\end{eqnarray}

Let $q_1$ (resp. $q_2$) denote the extinction probability for the
process starting with a single type-1 (resp. type-2) cell. The
following result is adapted from references \cite{Kimmel}, \cite{Jagers}, \cite{Haccou}:

\begin{theorem}
The probability of extinction of the process, $q=(q_1,q_2)$, is the solution of
equation
\begin{equation}
\label{extinction} 
(f_1(s,t),f_2(s,t))=(s,t) 
\end{equation}
that is closest to the origin in $[0,1]^2$.
\end{theorem}

\vspace{4 mm}

Note that $f_1(1,1)=f_2(1,1)=1$, so that $(1,1)$ is a solution
of (\ref{extinction}). Depending on the values of parameters
$\delta$, $\mu$ and $p$, it can happen: $i)$ $(1,1)$ is the only
solution of (\ref{extinction}) --and hence the extinction probability
is one--; $ii)$ there exists another solution of (\ref{extinction}) in $(0,1)^2$ --and non-extinction probability is positive.

\subsection{Critical surface}

In order to know which is the set of parameter values that make the multi-type branching process (and hence the antigenic variation model) become extinct with probability one, or survives forever with a positive probability, we look for the values $(s_0,t_0) \in [0,1]^2$  that satisfy:

\begin{eqnarray}
\label{extinction_2} 
\left\{ \delta + (1-\delta)[ \mu t_0 + (1-\mu) s_0 ] \right\}^2 &=&
s_0 \nonumber \\
(1-p) \sum_{k=1}^{\infty} {s_0}^{2^k} p^{k-1} &=& t_0.
\end{eqnarray}

Note that $s_0 \neq 1$ (resp. $s_0 \neq 0$) if and only if $t_0 \neq 1$ (resp. $t_0 \neq 0$). From these equations we get that the probability of non-extinction is positive if and only if there
exists $s_0 \in (0,1)$ such that:

\begin{equation}
\label{theequation} \left( \delta + (1-\delta)\left[ \mu (1-p)
\sum_{k=1}^{\infty} {s_0}^{2^k} p^{k-1} + (1-\mu) s_0 \right]
\right)^2 = s_0. \nonumber
\end{equation}

Let introduce the function:

\begin{equation}
g(s) = \left( \delta + (1-\delta)\left[ \mu (1-p)
\sum_{k=1}^{\infty} {s}^{2^k} p^{k-1} + (1-\mu) s \right]
\right)^2, \nonumber
\end{equation}

$s \in [0,1]$. Note that, for $\delta, \mu, p \in (0,1)$, $g(.)$ is a strictly increasing and convex function with $g(0)= \delta^2$ and $g(1)=1$. So that, there exists $s_0 \in (0,1)$ satisfying $g(s_0) = s_0$ if and only if $g'(1) > 1$.

It holds:

\begin{equation}
g'(1)= 2(1-\delta)\left[ 2 \mu (1-p)
\sum_{k=0}^{\infty} (2p)^{k} + (1-\mu) \right]. \nonumber
\end{equation}

{\bf Case 1:} $p \geq \frac{1}{2}$. $g'(1) = \infty$, and so there exists $s_0 \in (0, 1)$
such that $g(s_0) = s_0$. The probability of non-extinction is positive; we say the process lies in the {\em supercritical region}.

\vspace{4 mm}

{\bf Case 2:} $p < \frac{1}{2}$. We have 

\begin{equation}
g'(1) = \frac{2(1- \delta)}{1 - 2p}[1 + \mu - 2p]. \nonumber
\end{equation}

\vspace{4 mm}

{\bf Critical $\mu$:} For each pair $(\delta, p)$ we introduce the {\em critical value},
$\mu_c(\delta, p)$, defined by:

\begin{equation}
\frac{2(1- \delta)}{1 - 2p}[1 + \mu_c(\delta,p) - 2p] = 1, \nonumber
\end{equation}

$0 \leq \mu_c(\delta,p) \leq 1$.


It is clear that the probability of extinction is one --or strictly smaller than one-- depending on $\mu \leq \mu_c(\delta,p)$ --or $\mu > \mu_c(\delta,p)$.


We summarize the analysis in the following result:

\begin{theorem}
Let: 
\begin{equation}
\label{critical} 
\mu_c(\delta,p) = \left\{ 
\begin{array}{ll}
\min \left\{ 1, \max \left\{  0, \frac{(1-2p)}{2(1-\delta)}(2 \delta -1) \right\} \right\}  \;\;\; \mbox{if } p < \frac{1}{2} \\
0 \;\;\; \mbox{if } p \geq \frac{1}{2}
\end{array} 
\right.
\nonumber
\end{equation}
It holds, for each $(\delta,p,\mu)$:
\begin{itemize}
\item[i)] The process is {\em subcritical} (i.e. the extinction probability is one)
if $\mu \leq \mu_c(\delta,p)$.
\item[ii)] The process is {\em supercritical} (i.e. the probability of non-extinction
is strictly positive) if $\mu > \mu_c(\delta,p)$.
\end{itemize}
\end{theorem}

\begin{figure}
\includegraphics[width=0.5\textwidth]{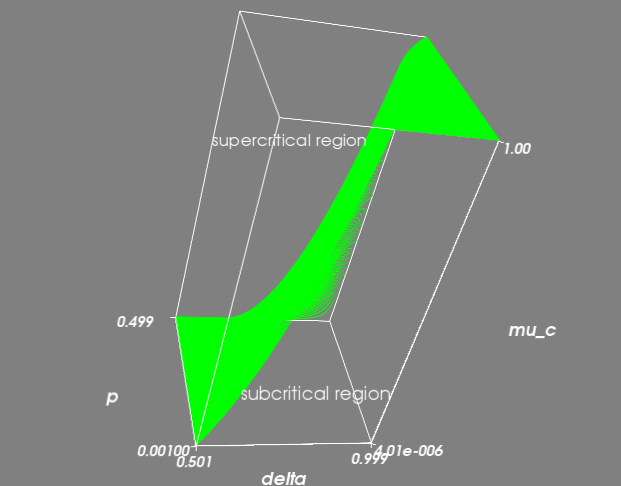}
\caption{Critical surface}
\label{fig2}
\end{figure}

{\bf Remark:} The critical surface does not depend on the initial condition of the process, provided that it starts with a finite number of type-1 and type-2 cells. That depends on the initial condition is the extinction probability in the superctitical region. 

\vspace{2 mm}

Figure \ref{fig2} shows the critical surface in the non-trivial region: $\delta > \frac{1}{2}$, $p < \frac{1}{2}$.

\begin{figure}
\includegraphics[width=0.5\textwidth]{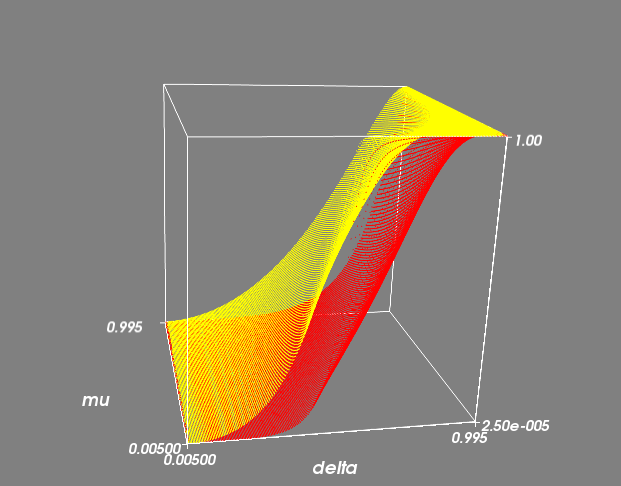}
\caption{Extinction probabilities for the process starting with one type-1 cell (yellow) and one type-2 cell (red). $p=0.65$.}
\label{fig3}
\end{figure}

Numerically solving equations (\ref{extinction_2}) in the supercritical region we obtain the extinction probability $q_1=s_0$ (resp. $q_2=t_0$) for the process starting with one type-1 cell (resp. one type-2 cell). In Figure \ref{fig3} we show these extinction probabilities computed for $p=0.65$. 

\section{Parasitemia Waves}

The model analyzed in the previous sections shows instability in the asymptotic limit $n \rightarrow \infty$, that means: the process becomes extinct (supposedly at an exponential or sub-exponential rate)  or it explodes exponentially fast (see for example ref. \cite{Kimmel}). All states except $Z_1=Z_2=0$ are transient and consequently does not exist any kind of {\em oscillatory regime} as has been experimentally observed in mammals infected with parasites or virus exhibiting antigenic variation (see ref. \cite{Ross}, \cite{Barry_Turner_1991}). In this section we introduce some modifications to obtain waves of parasitemia at least with high probability for a reasonable number of generations (let say, $n$ of the order of hundreds or thousands).

\subsection{Random parameters}

To avoid fast extinction or explosion in the number of cells we consider parameters $\mu$ and $\delta$ as random variables (the distribution of which depends on the state of the process at every $n$). 


In {\it Trypanosoma brucei} populations, several authors (\cite{Seed_Black}, \cite{Vassella}, \cite{Tyler}, \cite{Savill_Seed}) have proposed a mechanism that induces cell transformation from dividing to non-dividing forms in a density-dependent manner. This is a mechanism of self-regulation that prevents fast population explosions. Experimental results with immunosuppressed mice support this mechanism. Mathematical models involving a set of (deterministic, macroscopic) differential equations have been proposed to describe these concentration changes. 

We follow an analogous approach by updating the parameter $\delta$ along the evolution of the (microscopic) process. On the other hand, when the density of cells is small enough, we randomly update the value of $\mu$ to avoid the fast extinction of the process with high probability.  

\subsection{Varying $\delta$}

Here we consider parameter $\delta$ as a random variable updated with $n$. Fix $r_{\delta}>0$ and $0 \leq \delta_{min} \leq \delta_{max} \leq 1$.  We denote $R(n)$ the number of cells recognized by the immune system (that is, type-1 cells) at generation $n$. Let $X_{\delta}(n)$ be a random variable with Beta distribution of parameters $\alpha = \frac{R(n)}{r_{\delta}}$, $\beta = 1$.


We put:

\[ \delta(n) = \delta_{min} + (\delta_{max} - \delta_{min}) X_{\delta}(n)   , \,\, n=1,2,3,\ldots\]

and interpret $\delta_{min}$ as the background contribution of the immune system to the probability of death of sensitive cells; $\delta_{control}(n)=(\delta_{max} - \delta_{min}) X_{\delta}(n)$ takes into account the mechanism of self-control of parasite population at generation $n$: this mechanism regulates the rate of conversion of dividing cells into non-dividing forms.

\vspace{2 mm}

\begin{figure}
\vspace{-2 cm}
\includegraphics[width=0.6 \textwidth, angle=90]{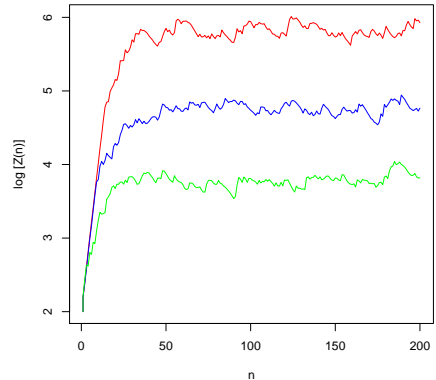}
\vspace{-2 cm}
\caption{Self-control: simulations of three Galton-Watson processes. $\delta_{min}=0.20$; $\delta_{max}=0.55$. $r_{\delta}=10^3$ (green); $r_{\delta}=10^4$ (blue); $r_{\delta}=10^5$ (red).}
\label{fig4}
\end{figure}

Figure \ref{fig4} shows the effect of $\delta_{control}$: the simulations correspond to realizations of Galton-Watson branching processes where cells at generation $n$ give birth to: 

\begin{itemize}
\item two cells with probability $(1-\delta(n))^2$, 
\item one cell with probability $2\delta(n)(1-\delta(n))$,
\item no cell with probability $\delta(n)^2$.   
\end{itemize}

$\delta_{min} = 0.20$, $\delta_{max} = 0.55$ and $r_{\delta}= 10^3, 10^4, 10^5$. As it can be seen, as $n$ grows up the total number of cells, $Z(n)$, fluctuates around a value not too far away from $r_{\delta}$.

\subsection{Varying $\mu$}

Rates of antigenic variation were measured experimentally in {\it Trypanosoma brucei} (see \cite{Barry_Turner_1991}, \cite{Turner_1997} and references therein). Values of parameter $\mu$ range from $10^{-3}$ to $10^{-5}-10^{-7}$ switches/cell/generation, depending on the conditions of infection (fly-transmitted or syringe-passaged). Low values of $\mu$ are comparable to mutation rates.


In our approach, we consider $\mu$ as a random variable updated with $n$:

\[ \mu(n) = \mu_{min} + (\mu_{max} - \mu_{min}) X_{\mu}(n), \,\, n=1,2,3,\ldots\]

where $0 \leq \mu_{min} \leq \mu_{max} \leq 1$ and for fixed $r_{\mu}>0$ $X_{\mu}(n)$ is a random variable with Beta distribution of parameters $\alpha = \frac{r_{\mu}}{R(n)}$, $\beta = 1$. As in the $\delta$ case, parameter $\alpha$ is updated with $n$. The effect of variable $\mu(n)$ is to favor the emergence of new variants when the population of cells is small enough.

\subsection{Simulations of parasitemia waves}

To visualize the emergence of parasitemia waves we implement simulations for the population of cells using the antigenic variation model with fixed $p$ and random variables $\delta(n)$ and $\mu(n)$ updated with $n$ as described before. Figure $5$ shows four typical realizations of the process for a given initial condition. Data are drawn every four generations. Black lines are used for the total population of cells, $Z(n)=R(n)+Q(n)$, and red lines for the population of resistant type-2 cells, $Q(n)$. 

\vspace{2 mm}

Values of parameters in simulations: 

\begin{itemize}
\item $p=0.65$
\item $\delta_{min}=0.60$, $\delta_{max}=0.95$
\item $\mu_{min}=0.004$, $\mu_{max}=0.01$
\item $r_{\delta}=10^4$, $r_{\mu}=10^2$
\item Initial conditions: $R(0)=5 \times 10^3$, $Q(0)=1$, $\delta_{initial}=0.60$, $\mu_{initial}=0.004$
\end{itemize}

We note that the vast majority of new variants have almost no effect at all. In figure (a) the process ended before $100$ generations, after a few parasitemia peaks at the beginning. Figure (b) shows several peaks of different heights. Figures (c) and (d) show extended regions with low concentration of parasites, after which the process emerges again. Because $p > \frac{1}{2}$ the process lies in the superctitical region for all times (whenever $Z(n)>0$). For each one of the independent processes starting with one type-1 cell or one type-2 cell at generation $n$, extinction probabilities computed from equations (\ref{extinction_2}) take values in the intervals: $s_0 \in (0.997, 1.000)$, $t_0 \in (0.935, 0.999)$; from those values it is possible to compute the total extinction probability for the process as a function of $n$.  

The examples show that our model is able to reproduce the essential aspects of the dynamics of antigenic variation. New elements can be easily incorporated to the model thanks to its simplicity.   

\begin{center}
\begin{table}[h]
\begin{tabular}{cc}
\includegraphics[height=40mm]{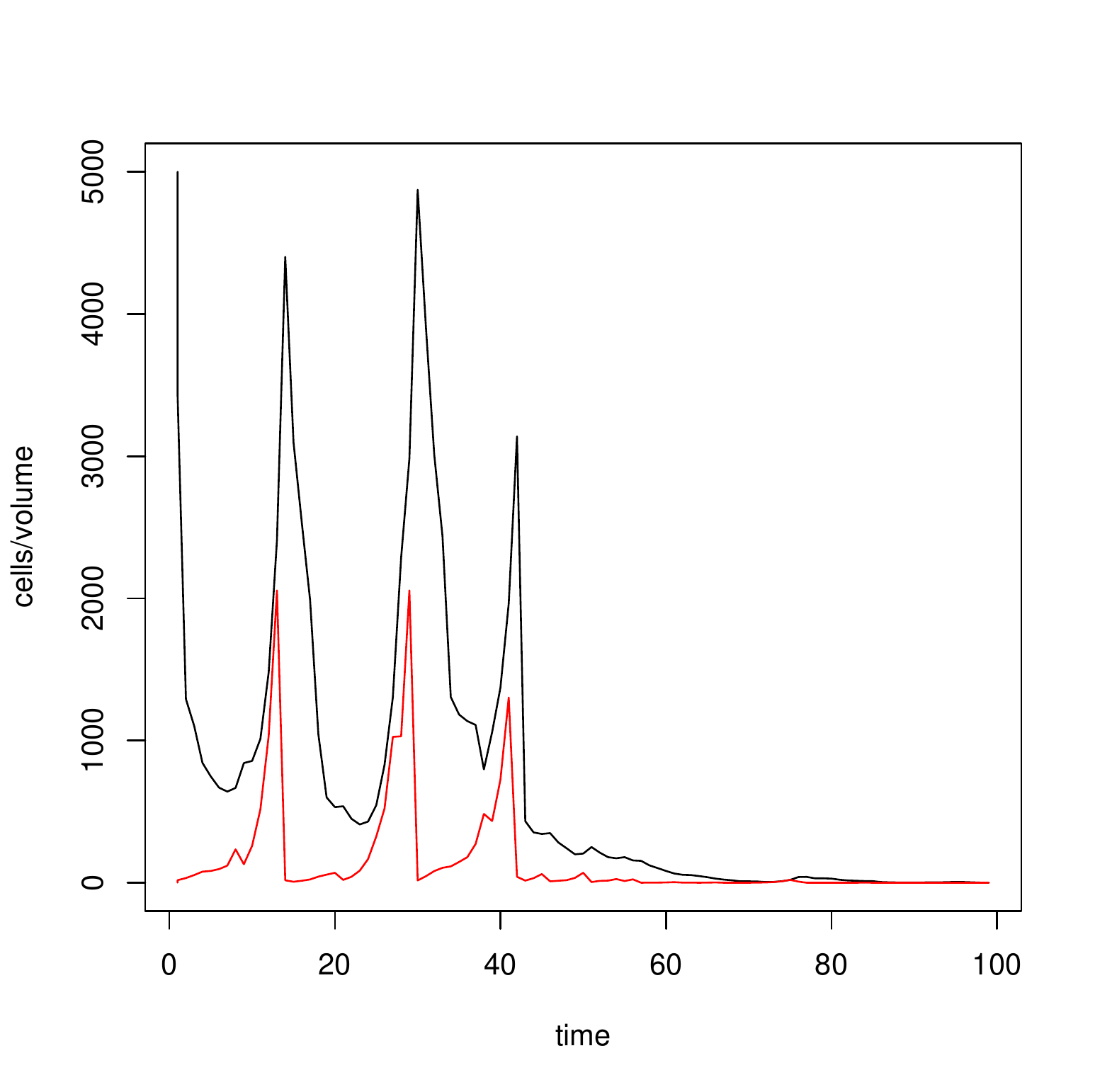} &
\includegraphics[height=40mm]{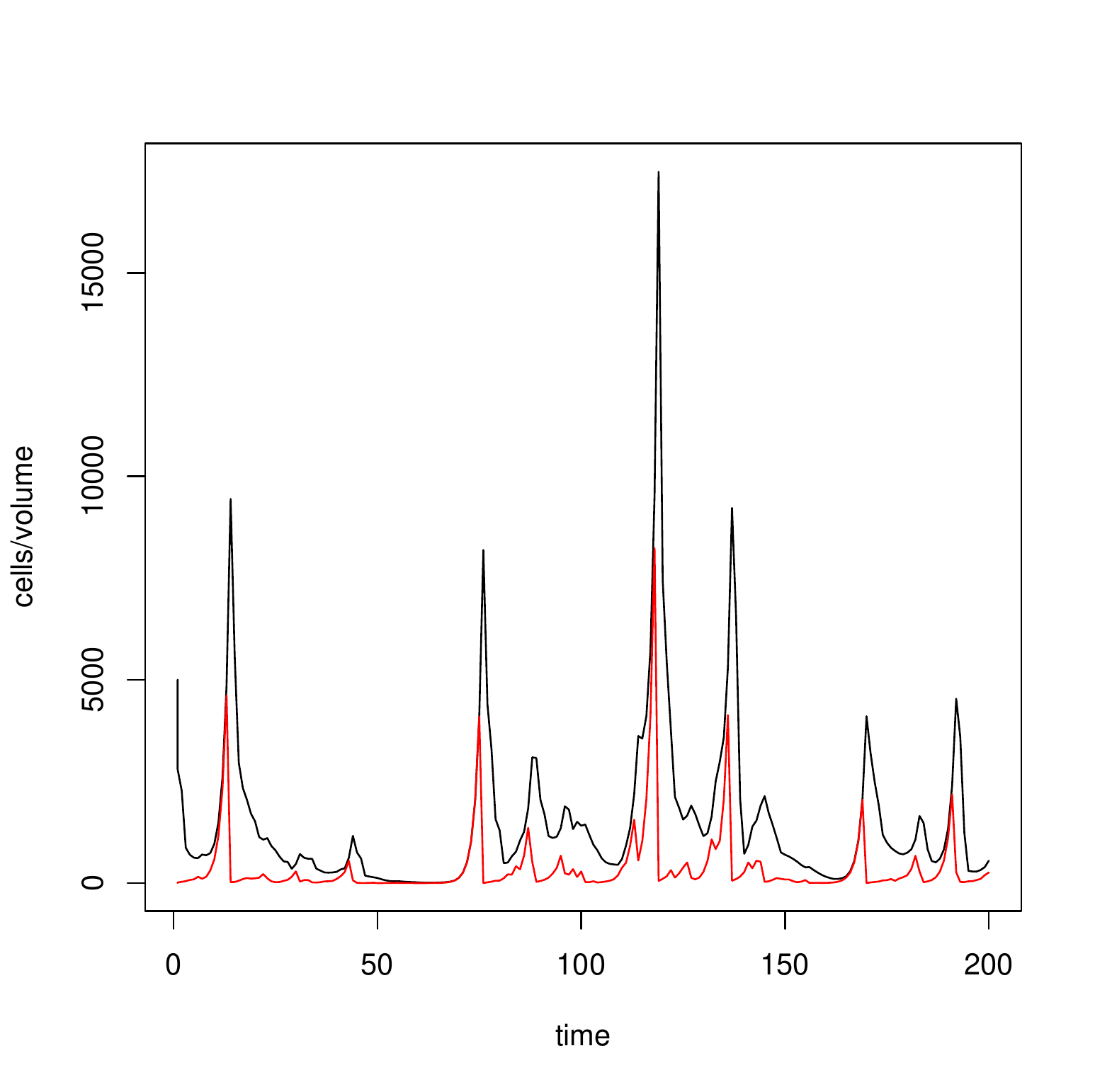} \\ 
FIG. 5: (a) & FIG. 5: (b) \\
\includegraphics[height=40mm]{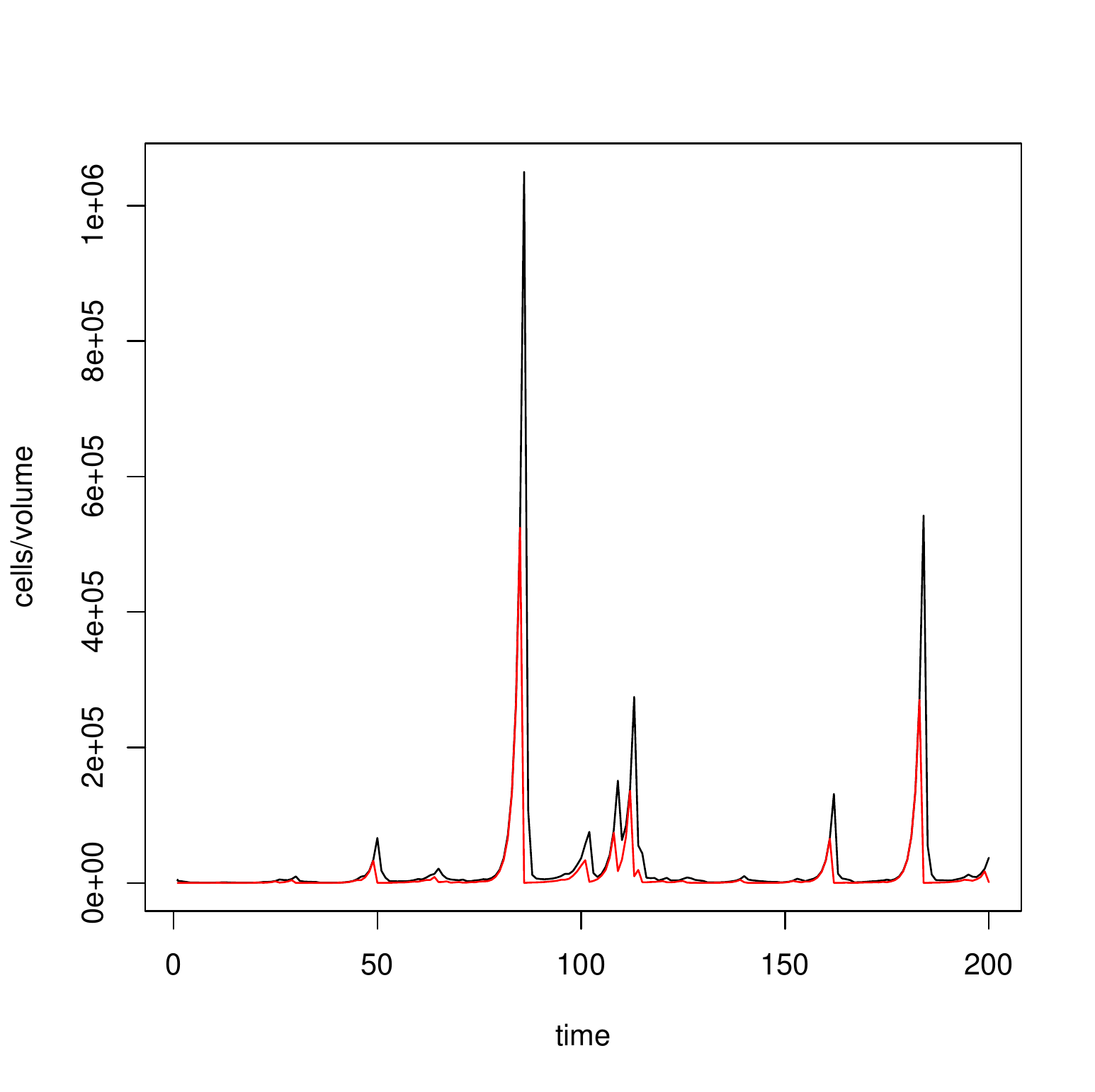} & 
\includegraphics[height=40mm]{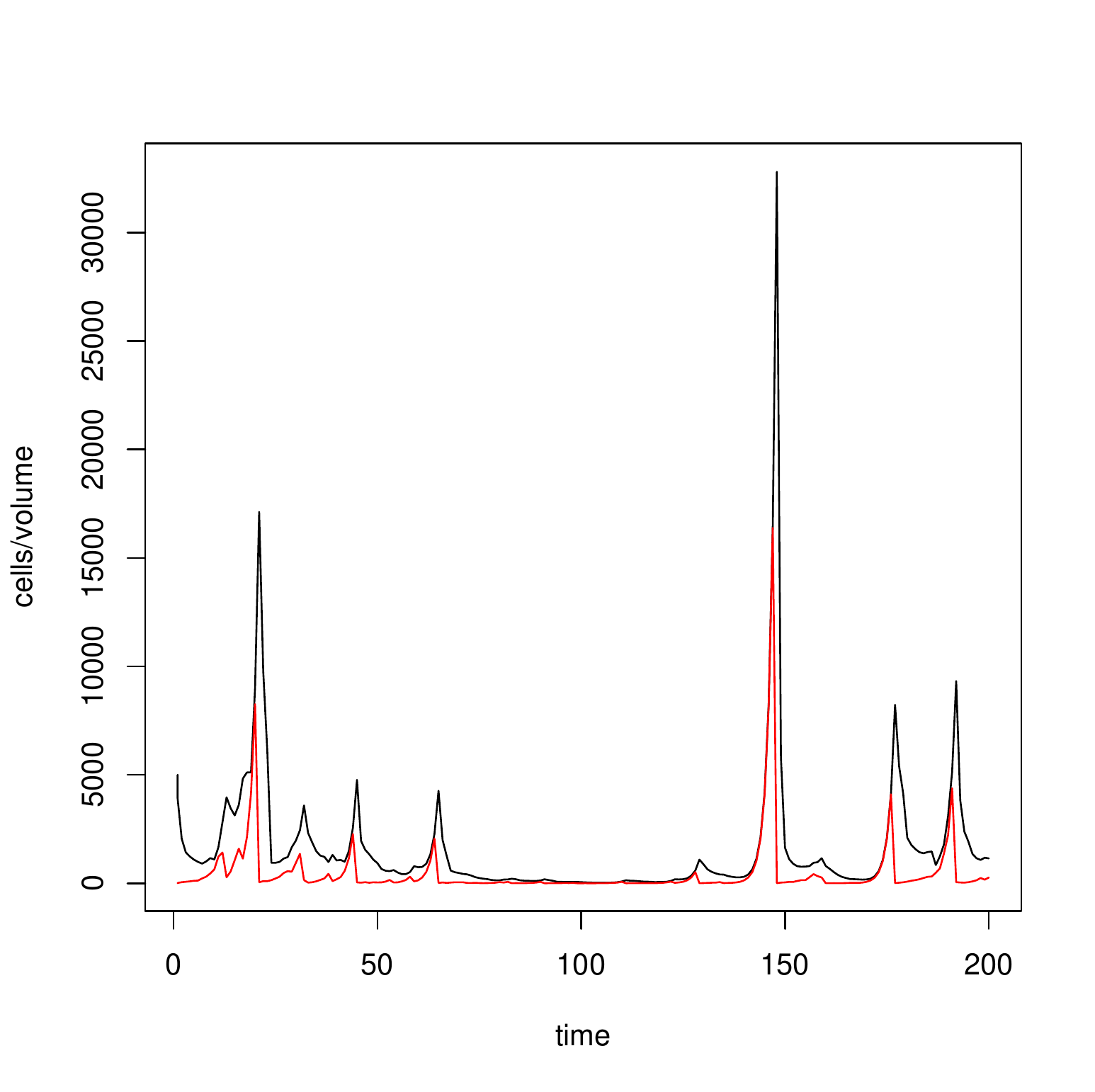} \\
FIG. 5: (c) & FIG. 5: (d)
\end{tabular}
\end{table}
\end{center}

\section{Conclusions}

In this paper we introduce a new approach to model the dynamics of antigenic variation using a multi-type branching process. The model considers the following aspects: efficiency of immune response against sensitive cells; rate at which new resistant variants appear; and a random modelling for the delay times spent by the immune system to recognize a new variant. 

We characterize analytically the subcritical and supercritical regions on the three-dimensional space of parameters. Simplicity and versatility are to be highlighted in our approach. By appropriately updating the set of parameters we introduce a density-dependent mechanism that accounts for qualitative behavior observed on in vivo experiments such as the characteristics peaks of parasitemia along an infection.

The model can be extended to include situations of higher complexity, as for instance: variable growth rates of the different antigenic variants (by adding random death parameters inside each clan of resistant cells), memory in the immune responses and delay times to recognize new antigen variants (by modifying the value of $p$ along the infection).   

In our opinion multi-type branching processes have a great potential to model evolution of host-parasite interactions like the particular case of antigenic variation. The approach presented here opens a variety of possibilities for future work.  

\vspace{4 mm}

\acknowledgements

This work has been partially supported by Agencia Nacional de Investigaci\'on e Innovaci\'on (ANII), Uruguay. We thank  Enrique Lessa, Sebasti\'an Castro and Guillermo Lamolle for helpful discussions.

\end{document}